# Probing Quantum Geometric Phases via Scanning Tunneling Microscopy

Chao Yan[1†], Mu-Wei Gao[1], Yue Zhao[1], Jia-Xin Yin[1,2†]

[1]Department of Physics, State Key Laboratory of Quantum Functional Materials, and Guangdong Basic Research Center of Excellence for Quantum Science, Southern University of Science and Technology, Shenzhen 518055, China.
[2]Quantum Science Center of Guangdong-Hong Kong-Macao Greater Bay Area, Shenzhen 518045, China

†Corresponding authors. E-mail: yanc@sustech.edu.cn; yinjx@sustech.edu.cn

**The quantum geometric phase intrinsically dictates the geometry, topology, and many-body correlations of electronic wave functions. While quantum geometric phases are conventionally inferred through momentum-space probes or macroscopic transport measurements, their direct visualization and quantification in real space have historically been restricted by the spatial averaging of bulk techniques. Scanning tunneling microscopy and spectroscopy (STM/STS) circumvent this limitation, leveraging atomic-scale spatial resolution and high energy sensitivity to resolve local electronic phase profiles directly. This review highlights recent progress across four representative methodologies: probing the Aharonov-Bohm (AB) geometric phase via nanoscale real space interferometry; extracting the Berry phase from defect-induced quasiparticle interference and wavefront dislocations; reconstructing the complex phase structure in symmetric systems, such as magic-angle graphene, using order parameter decomposition; and mapping the phase textures and topological defects of pair density wave (PDW) and charge density wave (CDW) in unconventional superconductors utilizing the numerical 2D lock-in technique. Together, these developments show how quantum phases can be translated onto real space and locally resolvable observables. Phase-resolved STM imaging provides stringent constraints on topological states of matter, symmetry-breaking patterns, and strong electronic correlations, outlining a robust framework for *in situ* phase engineering in quantum materials.**

## Introduction

The phase of a wave function represents a fundamental degree of freedom that dictates the symmetry,

topology, and coherence of electronic systems. Historically, condensed matter characterization focused primarily on amplitude-bound observables such as charge density and tunneling intensity, leaving the underlying phase information less accessible. In modern quantum materials research, however, the quantum phase is recognized as the central quantity driving non-local transport, stabilizing topological order, and mediating many-body correlations. Because the phase inherently encodes the global gauge structures and symmetry-breaking patterns of electronic systems, resolving and manipulating local phase textures in real space is essential for identifying emergent states of matter.

In 1981, G. Binnig and H. Rohrer successfully developed scanning tunneling microscopy (STM), exploiting the quantum tunneling effect between a metallic tip and a conducting sample surface[1-3]. STM enables simultaneous access to real-space atomic-scale surface morphology $T(\boldsymbol{r})$ and the local density of states (LDOS) $g(\boldsymbol{r},\mathrm{E})$ via differential conductance ($dI/dV$) mapping. Owing to its atomic-scale spatial resolution and exceptional sensitivity to local electronic structure, STM resolves spatial variations in electronic wave functions arising from quantum interference. This capability provides a powerful experimental platform for connecting quantum phase information with directly observable real-space quantities. Current STM-based quantum phase detection methodologies can be broadly categorized into four main approaches: constructing real-space interferometers to probe the AB effect; analyzing wavefront dislocations induced by quasiparticle scattering to resolve topological phases; utilizing order parameter decomposition in highly symmetric systems; and employing the numerical 2D lock-in technique to map the phase textures of strongly correlated orders, such as pair density wave (PDW).

**I. Probing the Aharonov-Bohm Effect via Nanoscale Interferometry**

In 1959, Aharonov and Bohm[4] proposed a gedanken experiment concerning the quantum interference of electrons moving in a magnetic field, known as the Aharonov-Bohm (AB) effect. The theory predicts that even when electrons propagate through regions with zero magnetic field, their wave-function phase can still be modified by the vector potential, resulting in observable changes in quantum interference patterns. Over the past several decades, experimental verification of the AB effect predominantly relied on measurements of conductance oscillations as a function of magnetic flux in ring-shaped conductors[5-15], which inherently lack spatial resolution and obscure microscopic interference behaviors. In 2009, theoretical work by Cano and Paul[16] demonstrated that on metal surfaces hosting a 2D electron gas (such as Cu (111) and Ag (111)), adatoms or structural defects can act as scattering centers for surface-state electrons (Fig. 1(a)). When electrons undergo multiple scattering events among these scatterers, they form closed real-space trajectories (Fig. 1(a)). In the absence of an external magnetic field, the electron wave functions propagate along two time-reversal-symmetric paths (clockwise and counterclockwise) and interfere, modulating the LDOS. Upon applying a perpendicular magnetic field, electrons traveling along these two paths accumulate different AB phases due to the vector potential. The resulting phase difference is given by $2\pi\Phi/\Phi_0$, where $\Phi$ is the magnetic flux enclosed by the trajectory and $\Phi_0$ is the flux quantum. Consequently, constructive interference in the LDOS evolves into an oscillatory behavior proportional to $\cos(2\pi\Phi/\Phi_0)$, manifesting as the magnetic field-dependent horizontal stripe patterns shown in Fig. 1(b). Therefore, by either fixing the STM tip position while varying the magnetic field, or fixing the magnetic field while scanning the STM tip, periodic AB oscillations can be directly detected in the conductance $dI/dV$. This theoretical work demonstrated that nanoscale AB interferometers could be realized via STM, providing an important conceptual foundation for later experimental studies. Following this proposal, a series of theoretical works further predicted that similar AB interference phenomena could be realized in a variety of systems, including strained graphene[17], topological insulators such as $Bi_2Se_3$[18,19], and configurations employing spin-polarized STM techniques[18,19].

A nanoscale AB interferometer was recently realized on a graphene/hexagonal boron nitride heterostructure[20]. Using STM tip-voltage pulses, Velasco and coworkers fabricated two circular p-type doped regions separated by approximately 150 nm. Due to charge accumulation, each region develops an electrostatic confinement potential, allowing it to act as an artificial atom[21-24]. By coupling these two quantum dots (Fig. 1(c)), an "artificial molecule" [20,25-27] hosting relativistic electron behavior was realized. By measuring the third-order differential conductance ($d^3I/dV^3$) under varying magnetic fields, they observed spatially resolved AB oscillations manifesting as alternating fringe patterns within a "figure-8" (region (iii)) shaped electron orbit. In this confined system, the electron wave function forms a closed "figure-8" trajectory, along which electrons propagate in opposite directions. When electrons return to the central region (iii), their wave functions acquire an AB phase due to the enclosed magnetic flux,

$$\Delta\phi = \frac{e}{h}\oint \boldsymbol{A}\cdot dl = \frac{e}{h}\Phi_B$$

where $\Phi_B$ denotes the magnetic flux threading the loop.

As the external magnetic field increases, the accumulated AB phase varies linearly, causing the LDOS to oscillate periodically between constructive interference and destructive interference, producing regular staggered pattern observed in Fig. 1(d). This work represents the first direct, spatially resolved visualization of AB oscillations in a coupled double-quantum-dot system in graphene, establishing a versatile platform for exploring geometric phases locally.

Subsequent studies have exploited small-angle twisted bilayer graphene structural reconstruction and its unique electronic properties to detect AB oscillations at the moiré scale[28], providing another route for probing geometric phase phenomena in complex device architectures.

**II. Probing the Berry Phase via Wavefront Dislocations**

In addition to the AB phase explicitly introduced by external magnetic fields, intrinsic topological defects within a material can be decoded under zero-field conditions through the analysis of wavefront dislocations induced by defect scattering. The Berry phase is a gauge-invariant geometric phase accumulated by the wave function along an adiabatic closed path in parameter space. In monolayer graphene, massless relativistic Dirac fermions traversing a closed orbit acquire a Berry phase of $\pi$[29-31], a feature that has been experimentally verified via the anomalous quantum Hall effect. Conventionally, measurements of the Berry phase rely on external electromagnetic fields to drive charged particles along closed trajectories encircling phase singularities, such as Dirac points[29,30,32-38]. This naturally raises the question of whether the Berry phase in graphene can be measured in the absence of an external magnetic field.

Realistic graphene samples invariably host atomic-scale defects that locally break sublattice symmetry and induce quasiparticle interference (QPI). For scattering events restricted to a single valley (namely intravalley scattering)[39-41], electron waves associated with the incoming and backscattered wavevectors $\boldsymbol{q}$ and $\boldsymbol{-q}$ possess pseudospins differing by $\pi$. This fundamental mismatch leads to destructive interference between the scattered electron waves, strongly suppressing intravalley scattering. In contrast, intervalley scattering[39-41] between valleys of opposite chiralities (K and K' valleys) is allowed, with the wave vectors connected by a polarization angle $\theta_q$. This process generates a prominent $\sqrt{3}\times\sqrt{3}\,R30^{\circ}$ charge density wave (CDW) modulation in real space[39]. Because intervalley scattering involves pseudospin rotation during wave-vector transfer, it directly reflects the valley topological properties of graphene. Thus, defect-induced scattering in graphene provides a real-space way to probe pseudospin winding and the associated Berry phase.

The wavefront dislocation is a topological wave phenomenon arising around phase singularities where the wave amplitude vanishes. As demonstrated by Nye and Berry[42], when a vortex wave interferes with a plane wave, the number of additional wavefronts appearing in real space is numerically equal to the angular momentum carried by the vortex wave. This principle enables the use of defect-induced changes in wavefront number to probe Berry phase physics in graphene under zero magnetic field.

In recent years, STM experiments have successfully observed quantum interference between pseudospin vortices and plane waves in graphene. In the vicinity of structural defects, quasiparticle scattering generates localized Friedel oscillations. Subsequent fast Fourier transform (FFT) analysis allows extraction of charge density oscillations along specific intervalley scattering directions, from which the number of wavefronts can be determined. Because the number of these additional wavefronts directly corresponds to the angular momentum of the pseudospin vortex, it provides access to the Berry phase. In monolayer graphene, when a single-atom defect introduces a locally rotationally symmetric potential, elastic scattering occurs between valleys of opposite chirality (K and K' valleys) (Fig. 2(b)). In this case, the pseudospin undergoes a rotation of $2\theta_q$, effectively generating a pseudospin vortex with angular momentum 2. This phenomenon was observed by isolating the intervalley components of hydrogen-adatom-induced Friedel oscillations (Fig. 2(c))[43] via fast Fourier transform (FFT) analysis. The appearance of two additional wavefronts along any scattering direction (Fig. 2(d)) confirmed a sublattice pseudospin winding number of 2, directly demonstrating the $\pi$ Berry phase under zero magnetic field. When a defect introduces a locally rotationally asymmetric potential, the scattered electron waves can form spiral configurations carrying specific angular momentum states and phase singularities, thereby generating new wavefront dislocations. Under an asymmetric defect potential, the wavefront structure around a single-atom defect was recently shown to evolve from two dislocations to a single dislocation (Fig. 2(e))[44], clarifying the interplay between orbital angular momentum coupling and pseudospin rotation in controlling phase singularities.

Expanding the focus from isolated phase singularities to collective electronic behaviors, several experiments have explored coherence and coupling between pseudospin vortices of different chiralities in monolayer graphene[45,46]. These scattering-induced wavefront dislocations remain topologically invariant under diverse symmetry-breaking perturbations that modify the pristine linear dispersion. Theoretical works predict that the characteristic pair of wavefront dislocations remains structurally stable even when the system is modified by an energy gap[47], magnetic doping[48], and trigonal-warped band structures[49]. These results offer direct guidelines for future zero-field scanning tunneling spectroscopy (STS) of perturbed Dirac fermions.

This real-space phase diagnostic has been successfully extended to more complex architectures. In Bernal stacked bilayer graphene (Fig. 2(f)), where the electronic bands and pseudospin configurations differ fundamentally from those of monolayer graphene[31,50-54], STM investigations demonstrated that intervalley scattering from single-atom defects (Fig. 2(g)) induces characteristic charge density oscillations[55]. While the number of extra wavefronts continues to map the pseudospin winding number, the exact defect site dictates the layer-specific redistribution of these topological dislocations (yielding 4, 2 or 0 additional wavefronts) (Fig. 2(h)), establishing a real-space framework for determining Berry phases in multilayer graphene systems. Furthermore, the correlated states emerging from flat bands in magic-angle graphene under different fillings have attracted intense interest in recent years[56-74]. Constructing accurate low-energy effective models underpins the description of these correlated phases, though their topological invariants remain critically sensitive to spatial symmetries. In the absence of decisive experimental criteria, identifying the appropriate low-energy Hamiltonian has historically been difficult. To resolve this ambiguity, theoretical calculations[75] demonstrated that introducing localized defects into distinct stacking regions of moiré superlattices generates scattering-induced LDOS features that distinguish between competing topological low-energy effective models (Fig. 2(i)), providing a real-space approach to model discrimination.

The above methods are primarily applicable to multivalley systems with weak spin-orbit coupling, such as graphene. For single-valley systems with strong spin-orbit coupling, including the surface states of 3D topological insulators, these approaches are no longer suitable. The quantum geometric tensor, whose symmetric part corresponds to the quantum metric and antisymmetric part to the Berry curvature, fully characterizes the geometric and topological properties of quantum states[76]. However, experimental extraction of the quantum geometric tensor has remained challenging. To address this issue, a scheme was proposed to measure the quantum geometric tensor utilizing spin-polarized STM[77]. By breaking time-reversal symmetry with an in-plane magnetic field to tilt the Dirac cone[78,79] (Fig. 2(j)) and measuring Friedel oscillations around magnetic impurities (Fig. 2(k)), momentum-resolved spin textures (Fig. 2(l)) can be extracted from the geometric amplitude of the oscillations. Since spin texture is directly related to the spinor phase of the eigenstates, the full quantum geometric tensor can in principle be reconstructed by taking the momentum derivative of the spin vector[80] (Fig. 2(m)). This approach is applicable not only to topological insulator surface states but also to systems with in-plane magnetic fields or energy gaps, providing a feasible route for direct measurements of the quantum geometric tensor in real space. In addition, a related theoretical proposal by Engström[81] suggests that impurity-induced quasiparticle interference detected by spin-polarized STM can probe the topological winding number of nodal points in 2D nodal superconductors.

In summary, the interplay between localized scattering and topological singularities enables the direct visualization of phase textures in low-dimensional quantum systems. The methodologies developed for graphene can be generalized to transition metal dichalcogenides (TMDCs)[82] and kagome lattices[83,84], where analogous spin or pseudospin textures encode rich geometric phases accessible via advanced scanning probe interference techniques.

## III. Investigating Phase Information via Order Parameter Decomposition

In magic-angle twisted bilayer graphene (MATBG), strong electron-electron correlations within the flat bands drive a rich phase diagram of insulating states[56,71,85], unconventional superconductivity[57,59,68], and topological magnetic states[60,73]. At the filling factor of $\nu = \pm 2$, distinguishing among various theoretically proposed energetically degenerate ground states[86-89] requires experimental techniques sensitive to the spatial symmetries of the underlying order parameters. With atomic-scale spatial resolution and exceptional energy sensitivity, STM/STS can distinguish competing phases by mapping the real-space signatures of symmetry breaking in the LDOS[85,90]. These real-space signatures of symmetry breaking establish decisive experimental constraints, elucidating the underlying microscopic mechanisms of correlated phases.

A prominent application of this real-space diagnostic is the identification of intervalley coherent (IVC) order in MATBG[91]. Spatially resolved tunneling spectroscopy reveals atomic-scale $\sqrt{3} \times \sqrt{3}\ R30°$ charge density modulations within both the correlated insulating ($\nu = \pm 2$) and superconducting regimes of magic-angle bilayer graphene, indicating the formation of a globally correlated state. Driven by intervalley electron scattering, these real-space $\sqrt{3} \times \sqrt{3}\ R30°$ modulations directly manifest IVC order. Fourier analysis of the spectroscopic maps across distinct stacking regions and filling factors (Fig. 3(a)) allows the identification of symmetry-breaking wave vectors, such as $\boldsymbol{Q}_{\mathrm{IVC}}$. By applying selective inverse FFT filtering, the real-space order parameter was reconstructed and decomposed into six independent complex order parameters adapted to the $C_{3v}$ point-group symmetry. Three of these components (IVC bond, IVC site A, and IVC site B) describe the translational and rotational symmetry of the valley-coherent modulation. Spatially mapping the amplitude and phase of each channel demonstrates that heterostrain dictates the macroscopic phase distribution (Fig. 3(b)). In weakly strained samples, the phases of the order parameters are spatially uniform, consistent with a time-reversal symmetric intervalley coherent (T-IVC) order. Conversely, in highly strained samples, the IVC phases exhibit stripe-like modulations, with a phase shift of $\Delta\theta = 2\pi/3$ between neighboring AA-stacked moiré regions. Subtracting the background phase gradient revealed a robust vortex-antivortex lattice, matching the theoretical predictions of the Kekulé spiral order.

This real-space geometric-phase diagnostics has been applied to monolayer graphene with hydrogen adatoms. Following the introduction of hydrogen adatoms, the scattering-induced Kekulé order parameter extracted from the STM image (Fig. 3(c)) can be decomposed into irreducible representations of bond-order amplitude (Fig. 3(d)) and bond order phase (Fig. 3(e))[92]. The bond-order phase accumulates a $2\pi$ winding around the impurity, forming a complete Kekulé phase vortex. Circling the defect with the STM tip in real space is mathematically equivalent to traversing a closed trajectory around the phase singularity (Dirac point) in momentum space. Consequently, the experimentally observed $2\pi$ Kekulé phase vortex provides direct evidence that monolayer graphene hosts a Berry phase of $\pi$. These studies underscore the power of symmetry-adapted order-parameter decomposition in decoding complex, many-body phase textures directly from STM data.

## IV. Quantum Phase Studies Based on 2D Lock-In Technique

In unconventional superconductors, coexisting and competing symmetry-breaking electronic orders (such as charge, spin, and pairing modulations) are closely intertwined. A prominent example is the PDW, an unusual superconducting state characterized by Cooper pairs carrying a finite center-of-mass momentum $\boldsymbol{Q}_{\mathrm{PDW}}$[93], which breaks long-range translational symmetry in the absence of an external magnetic field. In recent years, experimental signatures of PDW have been reported in a wide range of materials, including cuprates[94-99], iron-based superconductors[100-106], kagome systems[107-112], certain two-dimensional materials[113-116], and spin-triplet superconductors[117]. In addition to breaking U(1) gauge symmetry, PDW may also break translational or time-reversal symmetries, rendering its phase structure a key element for understanding its topology and microscopic origin. To isolate and precisely characterize the spatially modulated phase structures and topological defects of these intertwined orders, the numerical 2D lock-in technique has become essential.

Within the Ginzburg-Landau theoretical framework, the PDW order parameter can be expressed as[118]:

$$\Delta(\boldsymbol{r}) = \Delta_{\boldsymbol{Q}} e^{i\boldsymbol{Q}\cdot\boldsymbol{r}} + \Delta_{-\boldsymbol{Q}} e^{-i\boldsymbol{Q}\cdot\boldsymbol{r}},$$

where the phase factor $e^{i\boldsymbol{Q}\cdot\boldsymbol{r}}$ determines the spatial periodicity of the order parameter. Theoretical studies[118,119] have shown that the phase of a PDW is intrinsically linked to secondary orders, such as CDW ($\rho_Q$) or spin density wave (SDW, $S_Q$), via the relations:

$$\rho_Q \propto \Delta_Q \Delta^*_{-Q} + \text{c.c.} \qquad\qquad S_Q \propto i(\Delta_Q \Delta^*_{-Q} - \text{c.c.})$$

These relations indicate that the phases of CDW or SDW effectively encode the phase difference between PDW components. Therefore, by probing the phase of CDW or SDW, one can indirectly infer the phase structure of the underlying PDW, especially in cases where the PDW itself is difficult to image directly.

In 2008, Agterberg and Tsunetsugu theoretically predicted[119] that PDW can host fractional vortices (such as $\Phi_0/2$ or $\Phi_0/4$), which are accompanied by dislocations in the induced CDW or SDW. Specifically, a π phase shift across a half-dislocation in a PDW with wave vector $\boldsymbol{Q}$ nucleates a topological defect with a 2π phase winding in the induced CDW (2$\boldsymbol{Q}$) order.

To extract the spatial modulation of both amplitude and phase from STM images or spectroscopic maps, the 2D lock-in technique[97,100] is widely employed. For an arbitrary image $A(\boldsymbol{r}) = \sum_{\boldsymbol{Q}} a_{\boldsymbol{Q}}(\boldsymbol{r}) e^{i\boldsymbol{Q}\cdot\boldsymbol{r}}$, $\boldsymbol{Q}$ is the target wave vector and $a_Q(\boldsymbol{r})$ is the local complex amplitude associated with $\boldsymbol{Q}$ at the position $\boldsymbol{r}$, the complex amplitude in real space can be written as:

$$A_{\boldsymbol{Q}}(\boldsymbol{r}) = \int A(\boldsymbol{R}) e^{\mathrm{i}\boldsymbol{Q}\cdot\boldsymbol{R}} e^{-\frac{(\boldsymbol{r}-\boldsymbol{R})^2}{2\sigma_r^2}} \mathrm{d}\boldsymbol{R}$$

the complex amplitude in reciprocal space can be written as:

$$A_{\boldsymbol{Q}}(\boldsymbol{r}) = \mathcal{F}^{-1}\left(A_{\boldsymbol{Q}}(\boldsymbol{q})\right) = \mathcal{F}^{-1}\left[\mathcal{F}(A(\boldsymbol{r}) e^{\mathrm{i}\boldsymbol{Q}\cdot\boldsymbol{r}}) \cdot \frac{1}{\sqrt{2\pi}\sigma_q} e^{-\frac{\boldsymbol{q}^2}{2\sigma_q^2}}\right]$$

From the extracted complex amplitude $A_Q(\boldsymbol{r})$, the spatial amplitude and phase distributions are as follows:

$$|A_{\boldsymbol{Q}}(\boldsymbol{r})| = \sqrt{[\mathrm{Re}A_{\boldsymbol{Q}}(\boldsymbol{r})]^2 + [\mathrm{Im}A_{\boldsymbol{Q}}(\boldsymbol{r})]^2}$$

$$\Phi_{\boldsymbol{Q}}^A(\boldsymbol{r}) = \tan^{-1}\left(\frac{\mathrm{Im}A_{\boldsymbol{Q}}(\boldsymbol{r})}{\mathrm{Re}A_{\boldsymbol{Q}}(\boldsymbol{r})}\right)$$

Here, $\mathcal{F}$ denotes the Fourier transform, $\sigma_r$ and $\sigma_q$ represent the cutoff length in real space and the cutoff frequency in reciprocal space, respectively.

In practice, this lock-in procedure effectively filters the spatial maps by isolating electronic modulation components within the vicinity of the target wavevector $\boldsymbol{Q}$, thereby enabling direct visualization of the local phase textures and topological defects. For instance, applying Fourier transform to an energy gap map $\Delta(\boldsymbol{r})$ at a specific wavevector $\boldsymbol{Q}$ yields the filtered gap map $\Delta_{\boldsymbol{Q}}(\boldsymbol{r})$, along with its corresponding spatial magnitude $|\Delta_{\boldsymbol{Q}}(\boldsymbol{r})|$ and phase texture $\Phi_{\boldsymbol{Q}}^{\Delta}(\boldsymbol{r})$.

This spatial demodulation framework has been instrumental in identifying intertwined PDW orders in high $T_c$ cuprates. In $Bi_2Sr_2CaCu_2O_{8+\delta}$ (Bi-2212) system (Fig. 4(e)), spatial variations in the superconducting energy gap $\Delta(\boldsymbol{r})$ were tracked to resolve an $8a_0$ periodic gap modulation (Fig. 4(f)) coexisting with uniform superconductivity via STM[97]. Applying the 2D lock-in technique revealed that topological defects exhibiting a 2π phase winding (Fig. 4(g)) in the $4a_0$ CDW phase map spatially coincide with π phase jumps (Fig. 4(h)) in the $8a_0$ PDW phase texture. This observation confirmed the intertwined nature of the cuprate PDW state[118,119] and demonstrated that CDW emerges from the intertwining of PDW with superconductivity.

Analogous phenomena have also been explored in iron-based superconductors, which host multiple symmetry-breaking electronic orders such as nematic order[120]. A quasi-1D PDW state confined to domain walls was detected in epitaxial Fe(Te,Se)/$SrTiO_3$ thin films[100]. Spatially resolved tunneling spectroscopy mapped periodic modulations (Fig. 4(i)) in both the energy gap magnitude and coherence peak height at a characteristic wave vector $\boldsymbol{Q}_P \approx 0.28\boldsymbol{Q}_{\mathrm{Bragg}}^{\mathrm{Fe}}$. Fourier-filtered extraction via the 2D lock-in technique (Fig. 4(j)) successfully demonstrated that the predicted π-phase shifts in the PDW (Fig. 4(l)) order align with vortex-like 2π phase windings in the induced CDW (Fig. 4(k)), establishing the PDW as a primary electronic order pinned by local lattice distortions.

This phase-resolved methodology has further extended the frontier of PDW research into TMDCs. TMDCs host a rich variety of distinct electronic phases in the two-dimensional limit, ranging from conventional metals and semiconductors to highly correlated states such as CDW[113,121,122], valley electronics[82,123], Mott insulators[124-126], and superconductivity[113,124,127]. As a prototypical platform for investigating intertwined quantum orders, the single-crystal superconductor $NbSe_2$ simultaneously exhibits coexisting CDW, superconductivity, and PDW order.

To resolve these intertwined orders, Josephson scanning tunneling microscopy was employed to probe the coexisting CDW and superconductivity in $NbSe_2$[113]. Spatial maps of the zero-bias Josephson differential conductance unveiled a periodic modulation of the total pairing density $N_C(\boldsymbol{r})$ modulated by the host CDW lattice. Through Fourier transformation and the 2D lock-in technique, the pure PDW component $N_P(\boldsymbol{r})$ was successfully extracted from the measured pairing density image $N_C(\boldsymbol{r})$. The modulation wave vector $\boldsymbol{Q}_P$ of the PDW was found to be identical to the CDW wave vector $\boldsymbol{Q}_C$, while their phases exhibit a relative shift of 2π/3 in real space (Fig. 4(m)). These results demonstrate that the finite-momentum pairing (PDW) is intimately pinned by the primary density wave background, suggesting that the PDW arises from the intertwined quantum orders of CDW and superconductivity. Ginzburg-Landau analysis confirmed mutual suppression of the PDW and CDW order parameters at vortex cores, providing strong evidence that the PDW constitutes an independent quantum state. This spatial correlation provides the first experimental verification of a bulk independent PDW phase in TMDCs, offering a new perspective on intertwined quantum orders and unconventional superconducting mechanisms.

**Discussion and Outlook**

As an experimental platform that connects microscopic quantum phenomena with real-space observables, STM has undergone substantial methodological and technical advances in recent years. STM has evolved from a conventional tool for probing the LDOS into a core technique for investigating quantum phases. In this review, we have systematically surveyed STM/STS based progress in phase-resolved studies across several representative systems, including the AB geometric phase, the Berry phase in graphene, complex phases of many-body order parameters in magic-angle graphene, and the phase structure of PDW in unconventional superconductors.

From a methodological perspective, STM based quantum phase detection has converged toward four complementary research paradigms. First, by engineering closed electronic scattering trajectories, magnetoflux modulations of the LDOS expose the geometric AB phase. Second, intervalley scattering induced by defects and the resulting wavefront dislocations enable direct determination of the Berry phase in graphene under zero magnetic field conditions. Third, Fourier-based analysis combined with order-parameter decomposition allows reconstruction of complex order parameters in correlated systems, establishing real-space phase diagnostics for multivalley or multi-order intertwined systems. Fourth, the 2D lock-in technique maps local phase textures, revealing the topological defect structures of intertwined density waves. Together, these paradigms provide a rigorous framework for probing quantum phases locally, directionally, and with energy resolution.

Despite these advances, several important challenges and caveats merit attention. Because techniques like 2D lock-in filtering are numerical post-processing methods rather than direct raw observations, the resulting phase textures are sensitive to filtering parameters, window functions, and target wave vectors, necessitating caution when post-filtered signals approach background noise levels. Furthermore, in systems hosting strongly interwoven charge, spin, and pairing orders, developing decomposition schemes with stronger self-consistent physical constraints is vital for disentangling complex many-body phenomena.

Looking forward, continued development of STM techniques, such as ultra-low-temperature STM with higher energy resolution, spin-polarized STM, Josephson STM, time-resolved STM, and ultrafast STM will further expand the scope of quantum phase research. Utilizing time-resolved and terahertz-pulsed STM will enable real-time tracking of transient phase dynamics, including superconducting phase fluctuations, phase-slip events, and non-equilibrium phase transitions. Furthermore, localized electric and magnetic fields induced by the STM tip can be exploited to precisely engineer and manipulate phase textures *in situ* across 2D superconductors, topological boundary states, and moiré superlattices. Ultimately, the capacity to visualize, quantify, and manipulate quantum phases in real space will have profound implications for quantum phase engineering, topological information processing, and the development of next-generation quantum technologies.

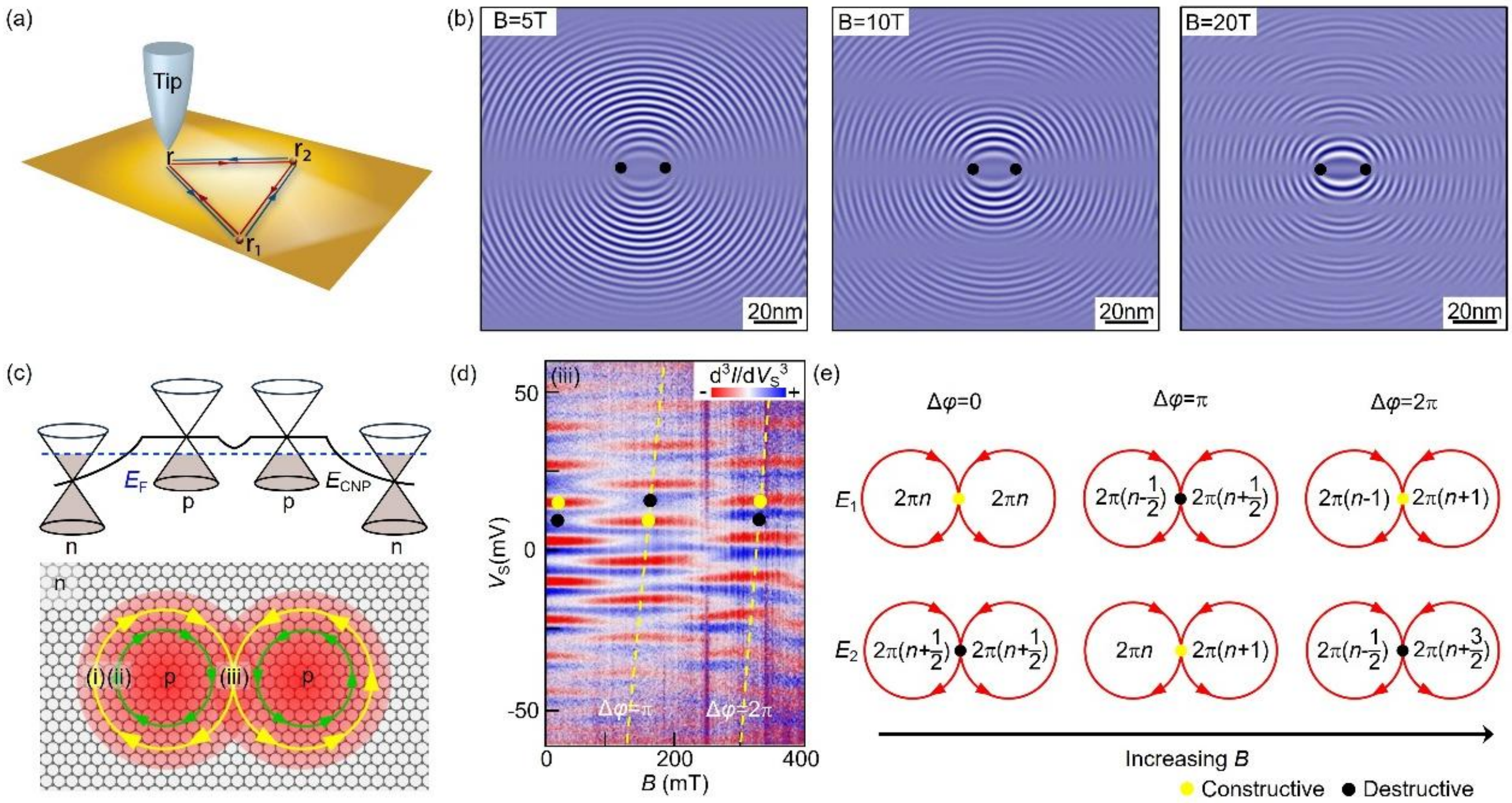

Fig. 1. Probing the AB effect using STM. (a) Schematic illustration of an STM interferometer, in which quantum interference arises from electron-wave scattering by two impurities located at $r_1$ and $r_2$, as reflected in the measured LDOS. Upon applying an external magnetic field, the AB effect modulates the interference process, resulting in periodic oscillations of the LDOS[16]; (b) Theoretical simulations of the interference patterns induced by two impurities on the Ag(111) surface under different magnetic fields, where the horizontal fringes originate from the AB effect[16]; (c) Schematic of a coupled double quantum dots structure, highlighting two types of electron trajectories: circular and "8" orbits[20]; (d) Magnetic-field evolution of the third-order differential conductance, measured at position (iii) in panel (c)[20]; (e) Quantum interference patterns at different energy levels at the center of the "8" orbit under an external magnetic field, where yellow and black dots denote constructive and destructive interference, respectively[20].

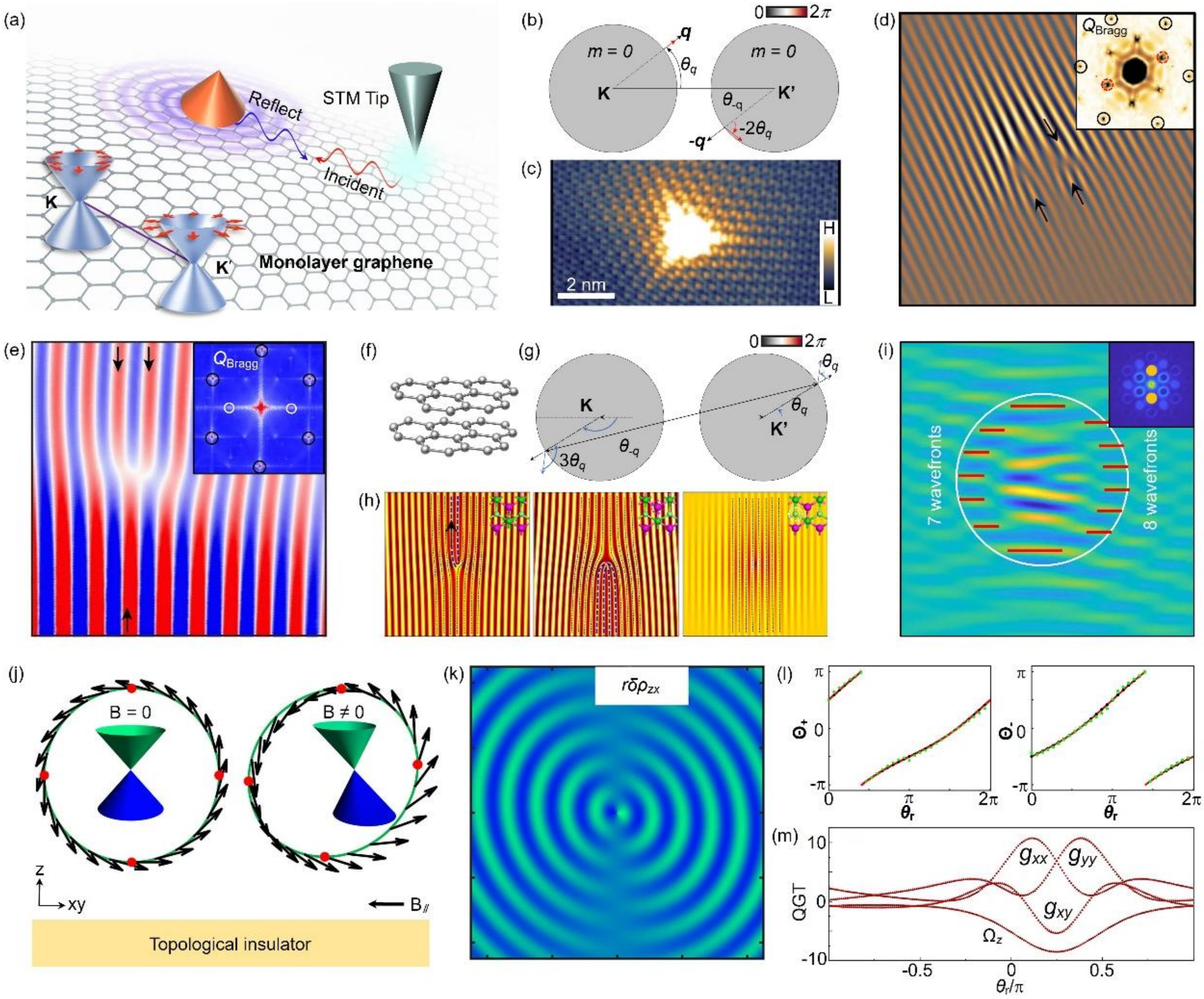

Fig. 2. Probing quantum phase via backscattering and wavefront dislocation. (a) Schematic of quantum interference between an STM tip and a defect; (b) Backscattering processes in monolayer graphene, where the wave vector carries no orbital angular momentum; intervalley scattering between the K and K' valleys induces a pseudospin rotation of $2\theta_q$[43]; (c) STM topographic image of monolayer graphene with hydrogen adatom[43]; (d) Inset shows FFT of panel (c). An inverse transform performed along the intervalley-scattering direction indicated by the red dashed circles yields the real-space charge-density oscillations, revealing two additional wavefront dislocations[43]; (e) Inset shows FFT image. An inverse transform along the intervalley-scattering directions marked by the white dashed circles produces real-space charge-density oscillations with one additional wavefront dislocation[44]; (f) Schematic of Bernal-stacked bilayer graphene; (g) Intervalley scattering processes in Bernal-stacked bilayer graphene; (h) Real-space charge-density oscillations along a specific intervalley-scattering direction in bilayer graphene induced by single-atom defects located at different lattice sites, producing 4, 2, and 0 additional wavefront dislocations, respectively[55]; (i) Calculated wavefronts in twisted bilayer graphene[75]; (j) Valley characteristics of surface states in a three-dimensional topological insulator with and without an in-plane magnetic field[77]; (k) Real-space oscillation patterns of the DOS measured by spin-polarized STM[77]; (l) Calculated spin textures[77]; (m) Calculated quantum metric and Berry curvature[80].

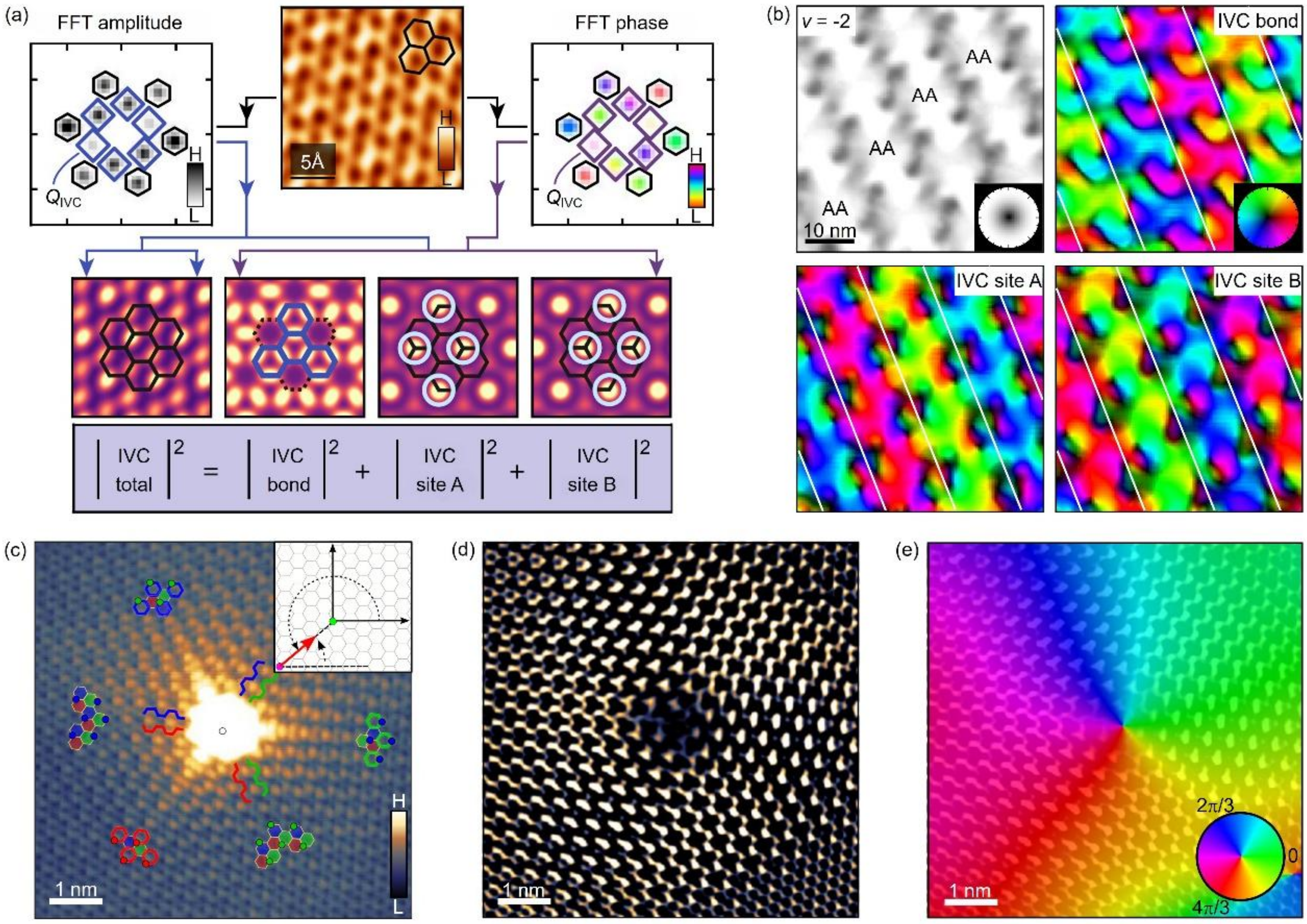

Fig. 3. Probing phase information via order-parameter decomposition. (a) Schematic illustration of decomposing a local order parameter based on FFT. The order parameter can be further resolved into three intervalley coherent components, whose amplitudes are combined along mutually orthogonal directions to form the total IVC order parameter[91]; (b) Spatial distribution of the valley-coherent order parameter in magic-angle twisted bilayer graphene at $v$=-2. The upper-left panel shows the global intensity of the valley-coherent wave function, while the IVC bond, IVC site A, and IVC site B maps reveal its spatial patterns; the corresponding phase information encodes the detailed shapes of these patterns[91]; (c) STM topographic image of graphene with hydrogen adatoms[92]; (d) Extracted Kekulé bond-order signal[92]; (e) Phase information of the Kekulé bond-order signal in panel (c)[92].

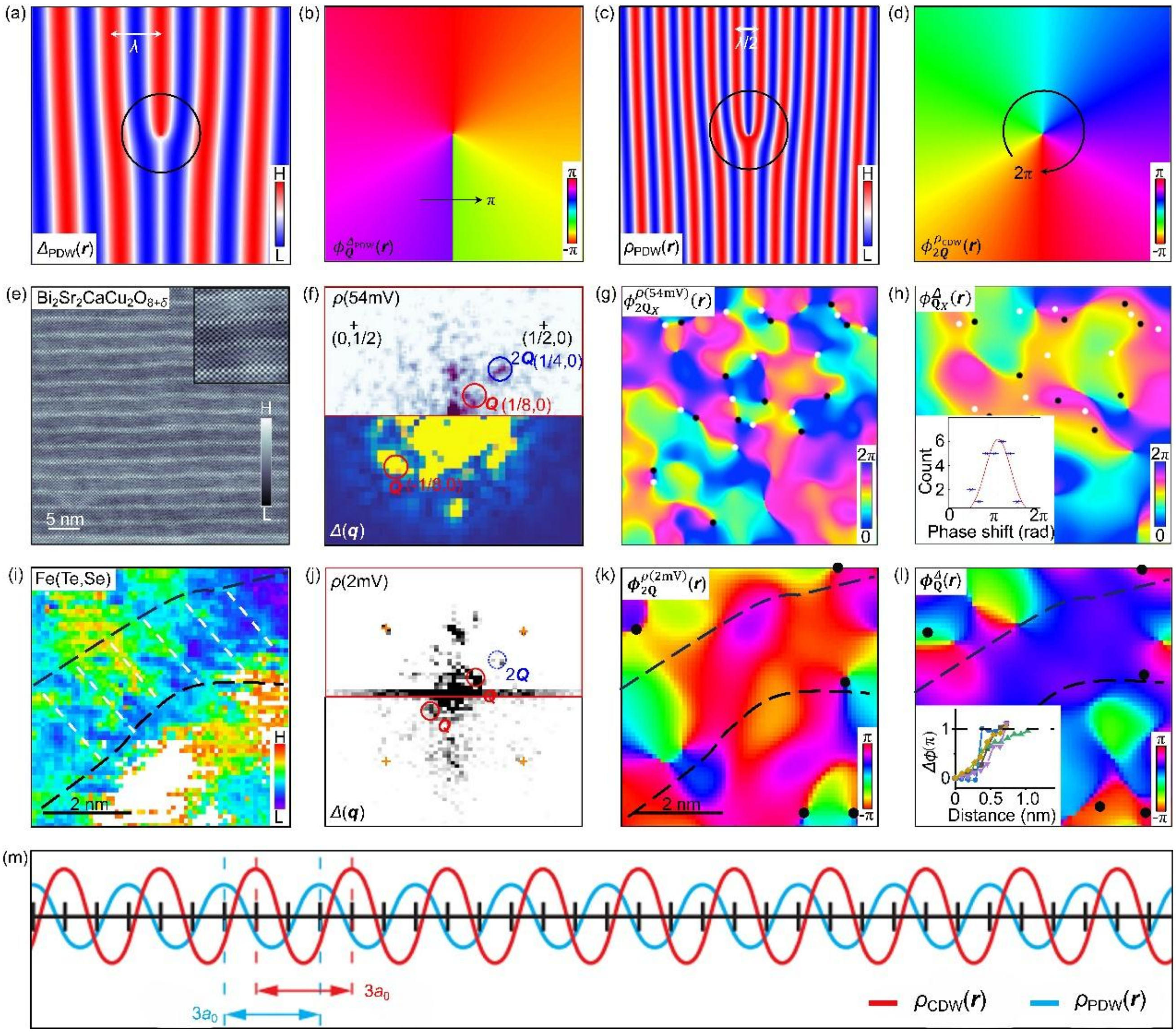

Fig. 4. Probing the phase of PDW and CDW using the 2D lock-in technique. (a) Schematic illustration of a half-vortex in a PDW[100]; (b) Corresponding phase map for panel (a)[100]; (c) Schematic illustration of a dislocation in a CDW[100]; (d) Corresponding phase map for panel (c)[100]; (e) STM topographic image of $Bi_2Sr_2CaCu_2O_{8+\delta}$[97]; (f) FFT at different energies, shown in the upper and lower panels, respectively[97]; (g) and (h) Phase distributions corresponding to the wave vectors $2\boldsymbol{Q}$ and $\boldsymbol{Q}$, respectively[97]; (i) Real-space distribution of the superconducting energy gap in Fe(Te,Se) thin films[100]; (j) FFT at different energies, shown in the upper and lower panels, respectively[100]; (k) and (l) Phase distributions corresponding to the wave vectors $2\boldsymbol{Q}$ and $\boldsymbol{Q}$, respectively[100]; (m) Phase distributions of the CDW and PDW in $NbSe_2$[113].